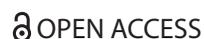

# Discretionary vs nondiscretionary in fiscal mechanism – non-automatic fiscal stabilisers vs automatic fiscal stabilisers

Vasile Brătian[a], Amelia Bucur[b], Camelia Oprean[a] and Cristina Tănăsescu[c]

[a]Faculty of Economics, Department of Finance-Accounting, "Lucian Blaga" University Sibiu, Sibiu, Romania; [b]Faculty of Sciences, Department of Mathematics, Informatics, "Lucian Blaga" University Sibiu, Sibiu, Romania; [c]Faculty of Economics, Department of Business Administration, "Lucian Blaga" University Sibiu, Sibiu, Romania

**ABSTRACT**

The goal of the present study is to increase the intelligibility of macroeconomic phenomena triggered by governmental intervention in economy by means of fiscal policies. During cyclical movements, fiscal policy can play an important role in order to help stabilise the economy. But discretionary policy usually implies implementation lags and is not automatically reversed when economic conditions change. In contrast, automatic fiscal stabilisers (SFA) ensure a prompter, and self-correcting fiscal response. The present study aims to tackle the topic of discretionary vs nondiscretionary characteristic of fiscal stabilisers (SF). In this context, the scope of the research undertaking is to launch a scientific debate over the definitions of the concepts of non-automatic fiscal stabilisers (SfnA) and SFAs. We describe how we can quantify the discretionary and non-discretionary character of the fiscal policy, by the analysis of the structure of the conventional budget balance (SBc), budget balance associated with the current GDP. In the final part of this article, we propose a quantitative equilibrium model for establishing the mathematical prerequisites for an SF to become automatic. Likewise, on the basis of the proposed mathematical model we have performed a qualitative analysis of the influence factors.

## 1. Introduction

In this section we will briefly discuss earlier work on the role of automatic fiscal stabilisers (SFA). During cyclical movements, fiscal policy can play an important role in order to help stabilise the economy. But discretionary policy usually implies implementation lags and is not automatically reversed when economic conditions change. In contrast, SFAs ensure a prompter, and self-correcting fiscal response. The SFA represents an institutional device which provides the non-discretionary character for the fiscal instruments which are operationalised through it (Dinga, 2009).

The academic literature does not clearly depict the distinction between discretionary fiscal policy and the fiscal policy operated through SFAs. This article makes precisely this conceptual clarification. In this matter, this study main contributions concern:

- the replacement of defining concepts by Aristotelian genus proximum and specific difference to their definition by listing sufficiency predicate (a minimum list of independent and mutual non-contradictory predicates, which qualifies the concepts covered);
- engendering a 'conceptual map' within the conceptual family of fiscal policy, which clearly and unequivocally grades SFAs within fiscal policy instruments (which the current literature does not do);
- the use of logical analysis (bivalent logic) in determining the sufficiency predicates, respectively in verbal formulation of definitions (which, again, the current literature does not do);
- establishing the mathematical conditions needed for an SF to be of automatic-type, by creating a quantitative equilibrium model.

The intended purpose of this research consists in conceptual and largely methodological clarification. The main goal of this study, through which the intended purpose is achieved, is to mutually clarify the two categories of fiscal policy fundamental instruments: explicit discretionary intervention, respectively implicit intervention by automatic stabilisers. The conceptual clarification of these two categories increases fiscal policy comprehensibility and, as a result, its efficacy, referred to in formulating the research intended purpose. Another goal of the present work is to provide the methodology for quantifying the SFAs under IMF and OECD definition.

## 2. Literature review

The post-2007 economic and financial crisis has reopened the debate on the effectiveness of fiscal policy as a tool of stabilisation of economic activity, including the relative merits of discretionary action versus automatic stabilisation. On one side of the debate, people have argued that discretionary fiscal policy is not an effective stabilisation tool. Especially from a political economy point of view, long decision and implementation lags associated with discretionary fiscal policy are often mentioned as arguments why such policies might be ineffective (Cogan, Cwik, Taylor, & Wieland, 2010). Others have argued the effectiveness of using automatic stabilisers only complemented by discretionary action by pointing out the presence of financially constrained households and accommodative monetary policy in crisis times (Christiano, Eichenbaum, & Rebelo, 2011; Coenen et al., 2012; Davig & Leeper, 2011).

Of the very few authors who considered the real importance of SFAs, Martner demonstrated in 2000 that the SFAs help to stimulate the economy in periods of recession and moderate it in booms, thus exercising a regulatory function, also showing that Governments have the option of allowing these automatic stabilisers to operate without intervention, or strengthening or restricting their effects through discretional polices (Martner, 2000).

The channels through which fiscal policy affects macroeconomic stability include supply-side effects of distortionary taxes, the procyclical behaviour of public spending induced by fiscal rules and the conventional effect of automatic stabilisers operating through disposable (permanent) income. Some authors (Andres & Domenech, 2006) investigate these channels and concludes that, contrary to what has been found in RBC models, distortionary taxes tend to reduce output volatility relative to lump-sum taxes when significant rigidities are present.

Another paper (Suescún, 2007) measures the size of automatic fiscal revenue stabilisers and evaluates their role in Latin America. It introduces a relatively rich tax structure into a dynamic, stochastic, multi-sector small open economy inhabited by rule-of-thumb consumers (who consume their wages and do not save or borrow) and Ricardian households to study the stabilising properties of different parameters of the tax code. The model captures many of the salient features of Latin America's business cycle facts and finds that the degree of smoothing provided by the automatic revenue stabilisers-described by various properties of the tax system-is negligible. Other authors analyse the effectiveness of the tax and transfer systems in the EU and the US to provide income insurance through automatic stabilisation in the recent economic crisis (Dolls, Fuest, & Peichl, 2012). Actually, the post-2007 economic and financial crisis has reopened the debate on the effectiveness of fiscal policy as a tool of stabilisation of economic activity, including the relative merits of discretionary action versus automatic stabilisation (Veld, Larch, & Vandeweyer, 2013).

The scientific literature provides quantitative estimates of the effects of the automatic stabilisers, but we mention that the SFAs are most commonly estimated with the elasticity approach (Fedelino, Ivanova, & Horton, 2009). The degree of output smoothing is estimated also by the use of simulation models (Dolls et al., 2012; Follette & Lutz, 2010).

## 3. Discretionary vs non-discretionary in fiscal mechanism – conceptual aspects

In what follows we will present the concepts and what we consider about the topic addressed.

Public policy (PP) is the product of public organisation (the output of the organisation/institution). By PP we understand a policy which verifies the following sufficient predicates (Dinga, 2009, p. 77):

- 'it is instrumented by a public organisation/institution licenced by law (usually through the fundamental legal law – the Constitution);
- it targets a public objective (or a coherent set of objectives);
- the targeted public objective pursued is relevant at both the macro-economic and macro-social level;
- it has a permanent and continuous character;
- it contains a computational mechanism (algorithmic-type) that describes the input-output relationship (or, more generally, the cause-effect relationship), which accompanies its instrumentalisation;
- there is a procedure associated to the computational mechanism which starts this mechanism.'

According to the causal criterion and formal criterion, we have the following categories of PP (Dinga, 2009, p.78):

  - according to the causal criterion, the PP may be of two categories:
  1. direct PP is the policy that determines the variation of the target variable directly without mediation of a variable (e.g., the administrative setting of a price for a particular good or service);

2. indirect PP is the policy that determines the variation of the target variable indirectly through the intermediation of a variable (e.g., variation in tax base through tax-rate variation).

- according to the formal criterion, PP may be of two categories:

1. discretionary PP (explicit PP) is that PP which obtains the variation of the target variable by a formal action (following a decision) of the organisation/institution responsible for the specific PP; [1]
2. non-Discretionary PP (implicit PP) is that PP which obtains the variation of the target variable without an action (following a decision) of the organisation/institution responsible for the specific PP.

*Fiscal policy* is a PP, also called Public Policy of adjustment (PPA) in which we will not find the direct character of the public intervention, representing an assembly of norms, institutions and procedures aimed to administer, from the perspective of the public authorities, the macroeconomic equilibrium in the real economy, the control of taxation rates and governmental expenditure (Dinga, 2009, p. 84).

Taking account of the above-mentioned criteria, with regard to the PP (causal criterion and the formal criterion), and taking into account the indirect nature of the fiscal policy (PPA), it follows that there are logically possible two categories fiscal policy:

1. indirect explicit fiscal policy (discretionary);
2. indirect implicit fiscal policy (non-discretionary).

*The fiscal mechanism* is a species of the fiscal system, representing a set of fiscal methods, techniques and tools by the use of which it provides:

determination, disposal and collection of taxes, fees, contributions and other amounts owed to the consolidated budget of the State;
determination and allocation of budgetary expenditure.

In accordance with the differences made above, concerning the categories of fiscal policy, we have two categories of fiscal mechanisms (Dinga, 2011, p.117):

1. discretionary fiscal mechanism, respectively that mechanism indirectly causative generated and realised by formal explicit actions of design, implementation (functioning) and monitoring of fiscal policy or fiscal instruments. Discretionary fiscal mechanism is based on the explicit fiscal instruments[2];
2. non-discretionary fiscal mechanism, respectively that mechanism indirectly causative generated and realised by formal implicit actions of design, implementation (functioning) and monitoring of fiscal policy or fiscal instruments. Non-discretionary fiscal mechanism is based on SFAs.[3]

*Explicit fiscal instruments* are those instruments by which we obtain the variation of the target variable through an action that is the result of an explicit formal decision of the public organisation responsible for the fiscal policy – Government or, in particular, the Ministry of Public Finances).

*Implicit fiscal instruments* are those instruments by which we obtain the variation of the target variable without an action that is the result of an explicit formal decision of the public organisation responsible for the fiscal policy – Government or, in particular, the Ministry of Public Finances).

The aim of the explicit instruments (discretionary), related to the fiscal mechanism, can be both anti-cyclical (especially on the decrease), and pro-cyclical (especially on the increase).

The aim of the implicit instruments (non-discretionary), related to the fiscal mechanism is to reduce the volatility of the macroeconomic output (GDP).

By reducing the volatility of GDP, it means that potential GDP is exceeded by current GDP – on the increase (current GDP > potential GDP) or current GDP is decreased under potential GDP – on the decrease (current GDP < potential GDP) (Eichengreen, 1997).

## 4. Quantification of discretionary and non-discretionary character in fiscal policy

In the following we describe how we can quantify the discretionary and non-discretionary character of the fiscal policy.

Discretionary character and non-discretionary character of a fiscal policy, with its mechanisms and instruments, can be quantified by the analysis of the structure of the conventional budget balance (SBc), budget balance associated with the current GDP.

According to the IMF and the OECD methodology, the structure of the SBC is made up of structural budget balance (SBS)[4] – also called *cyclically-adjusted balance*[5] – to which is added the cyclical budget balance (SBc) – also called *cyclical balance* – and it is given by the following relationship:

$$SBC = SBS \cup SBc \quad (1)$$

or

$$SBC = SBS + SBc \quad (2)$$

According to the Maastricht Treaty and the Stability and Growth Pact (Fiscal Pact [FP]), the maximum ceiling permitted for national public deficit is 3% of GDP at market prices. Also, the FP provides an SBS up to 0,5% of current GDP or max 1% if the ratio between public debt and GDP is significantly below 60% of GDP and the risks, in terms of the long-term sustainability of public finances, are low (Treaty on stability, coordination & governance, European Council (TSCG – UEM/ro), 2012, pp. 11–12).

The SBc is the result of two types of mechanisms that entail it:

1. discretionary mechanism, mechanism which generates the SBS, with discretionary instruments;
2. non-discretionary mechanism, mechanism which generates the SBc, with non-discretionary instruments.

Quantitative change (modification) of the SBC (ΔSBc) can be broken down into:

1. quantitative change of the SBS (ΔSBS), in response to the discretionary instruments;
2. quantitative change of SBC (ΔSBc), in response to the non-discretionary instruments.

According to the above, we can write the following in sequence:

$$\Delta SBC = \Delta SBS + \Delta SBc$$
$$where \quad (3)$$
$$\Delta SBc = SFA = \Delta SBC - \Delta SBS$$

where SFA represent automatic fiscal stabilisers.

Within the European Union there are proposed and used two methods of quantifying the SBS, i.e. the discretionary component of the SBC.

These methods are:

aggregate method, proposed by the International Monetary Fund (IMF methodology); disaggregate method, proposed by the European Commission (OECD methodology).

Regarding the aggregate method of quantifying the SBS (Fedelino, Ivanova, Horton, 2009), quantitative change of the SBS (modification of the cyclically-adjusted balance), comes from the budgetary cyclically-adjusted revenue and expenditure.

The component of the cyclically-adjusted revenue ($V_{ac}$) is defined as follows:

$$V_{ac} = V \left( \frac{Yp}{Y} \right)^{\varepsilon_V} \quad (4)$$

where:
V = nominal budgetary revenue (current);
Yp = potential GDP;
Y = current GDP;
$\varepsilon_V$ = revenue elasticity depending on the output gap.

Gap is determined as follows:

$$gap = \left( \frac{Y - Yp}{Yp} \right) \quad (5)$$

The component of the cyclically-adjusted expenditure ($C_{ac}$) is defined as follows:

$$C_{ac} = C \left( \frac{Yp}{Y} \right)^{\varepsilon_C} \quad (6)$$

where:
C = nominal budget expenditure (current);
$\varepsilon_C$ = expenditure elasticity depending on the gap.

As a result of the above, the SBS[6] is:

$$SBS = V \left( \frac{Yp}{Y} \right)^{\varepsilon_V} - C \left( \frac{Yp}{Y} \right)^{\varepsilon_C} \quad (7)$$

Through simplification, assuming $\varepsilon_V = 1$ (i.e., revenues are perfectly correlated with the cycle) and $\varepsilon_C = 0$ (i.e., expenditure are not affected by cycle), the structural budget balance *(SBS)* becomes: $SBS = V \left( \frac{Yp}{Y} \right) - C$, and the cyclical budget balance *(SBc)* is:

$$SBc = SBC - SBS = V\left(1 - \frac{Yp}{Y}\right) - C = \frac{V}{Y} \cdot Yp \cdot gap \qquad (8)$$

*Observation:*

There are four types of public revenue and one type of public expenditure, sensitive to the cycle.

In terms of budgetary revenue types, sensitive to the cycle, there are distinguished: a) personal income tax; b) corporate income tax); c) contributions for social assistance; parafiscal levies.

*In terms of the type of budgetary expenditure is distinguished: unemployment indemnity expenditure.*

Regarding the disaggregate method of quantifying the SBS (Girouard & Andre, 2005), according to this methodology, the SBS, i.e. the cyclically adjusted balance is:

$$SBS = \frac{\left[\left(\sum_{i=1}^{4} T_i^*\right) - C^* + X\right]}{Yp} \qquad (9)$$

where:

$C^*$ = cyclically adjusted budgetary expenditure;
$T_i^*$ = cyclically-adjusted budgetary revenue, by category ($i = \overline{1,4}$);
X = non-fiscale budgetary revenue - capital expenditure net of interest;
Yp = potential GDP.

The component of the cyclically-adjusted revenue is defined as follows:

$$\frac{T_i^*}{T_i} = \left(\frac{Yp}{Y}\right)^{\varepsilon_{t_i,y}} \qquad (10)$$

where:

$T_i$ = current budgetary revenue by category;
Y = current GDP;
$\varepsilon_{ti,y}$ = budgetary revenue elasticity, by category, depending on the output gap.

The component of the cyclically-adjusted expenditure is defined as follows:

$$\frac{C^*}{C_i} = \left(\frac{U^*}{U}\right)^{\varepsilon_{c,u}} \qquad (11)$$

where:

$C_i$ = current budgetary expenditure;
U = current level of unemployment;
$U^*$ = structural level of unemployment;
$\varepsilon_{g,u}$ = budgetary expenditure elasticity depending on the ratio between the structural level and the current level of unemployment.

Taking into account the above relationships, the SBS can be derived as follows:

$$SBS = \frac{\left\{\left[\sum_{i=1}^{4} T_i \left(\frac{Yp}{Y}\right)^{\varepsilon_{t_i,y}} - C\left(\frac{U^*}{U}\right)^{\varepsilon_{c,u}} + X\right]\right\}}{Yp} \quad (12)$$

The above methodologies quantify the result generated by the discretionary fiscal instruments (the result of SBS) and non-discretionary fiscal instruments (the result of SBc) over the SBc.

These methodologies use indicators which reflect the degree of utilisation of resources (deviation between the current GDP and potential GDP and deviation between actual unemployment and structural unemployment). The calculations regarding the estimation of potential GDP and structural unemployment are subject to measurement errors. Estimate of these indicators is an approximation because they do not take into account the driving forces of the business cycle, which change over time, with implications on budgetary revenue and expenditure.

Along with the above methods, the following may be mentioned:

quantification method based on the use of regression (Blanchard, 1990);
quantification method based on VAR methodology (Dalsgaard & De Serres, 1999);
quantification method based on models with undetectable components (Camba-Mandez & Lamo, 2002).

## 5. Discretionary vs non-discretionary in fiscal mechanism – *non-automatic fiscal stabilisers vs automatic fiscal stabilisers*

In the following section we will discuss SFs, a component on which the fiscal mechanism is based.

In this regard we proceed by completing the following steps:[7]

- we will define *the category* [8] called stabiliser (S);
- we will define *the order* called the macroeconomic stabiliser *(SM)*;
- we will define the family called the fiscal stabiliser (SF);[9]
- we will define the genus of non-automatic fiscal stabiliser (SFnA);
- we will define the species of non-automatic fiscal stabiliser at the level of public revenue (SFnAv);
- we will define the species of non-automatic fiscal stabiliser at the level of public expenditure (SFnAc);
- we will define the genus of automatic fiscal stabiliser (SFA);
- we will define the species of automatic fiscal stabiliser at the level of public revenue (SFAv);
- we will define the species of automatic fiscal stabiliser at the level of public expenditure (SFAc*)*;
- we will *formulate* conclusions.

### 5.1. Definition of the stabiliser (S)

In order to obtain the definition of stabiliser will try to identify the sufficient predicates. In our opinion, the sufficient predicates of the stabiliser are (Brătian, 2014):

1. it is an instrument, in the form of an institutional-type[10] device[11], by means of which it controls the change[12];
2. its action aims to reduce the gap between the actual change and desired change[13];
3. its action is opposite to the change;
4. its action is overproportionally in relation to the change.

Therefore, we may draw the following concluding remarks regarding the definition of the stabiliser, as a phenomenological characteristic: the (S) stabiliser is an instrument represented by a (normative) institutional device enabling the control of change; its actions are meant to decrease the discrepancy between the real and desired change, as it counters change and it is also overproportional compared to the change.

### 5.2. Definition of the macroeconomic stabiliser

In our opinion, the sufficient predicates of the SM are:

1. it is a stabiliser (S);
2. controls the quantitative change of the macroeconomic output (GDP);
3. it aims to reduce the volatility of the macroeconomic output (GDP)[14].

Therefore, we may draw the following concluding remarks regarding the definition of the SM, as a phenomenological characteristic: the SM is a stabiliser that controls the quantitative change of the macroeconomic output and it aims to reduce the volatility of the macroeconomic output.

### 5.3. Definition of the fiscal stabiliser

In our opinion, the sufficient predicates of the SF are:

1. it is an SM;
2. its action is formal normative, i.e., by this device, the aim is achieved through a formal normative action.

Therefore, we may draw the following concluding remarks regarding the definition of the SF, as a phenomenological characteristic: the SF is an SM and its action is formal normative, i.e., by this device, the aim is achieved through a formal normative action.

### 5.4. Definition of the non-automatic fiscal stabiliser

In our opinion, the SFnA

1. it is an SF;
2. its action is formal explicit, i.e., by this device, the aim is achieved through a discretionary, formal explicit action.

Therefore, we may draw the following concluding remarks regarding the definition of the SfnA, as a phenomenological characteristic: the SFnA is an SF and its action is formal explicit, i.e., by this device, the aim is achieved through a discretionary, formal explicit action.

### *5.5. Definition of the non-automatic fiscal stabiliser on budgetary revenue*

In our opinion, the SFnAv on budgetary revenue is generated by the simultaneous checking of three sufficient predicates:

1. it is an SFnA;
2. it controles in a linear way the quantitative change of the budgetary revenue;
3. its action is discrete, i.e., by this device, the aim is achieved through a discrete action.

Therefore, we may draw the following concluding remarks regarding the definition of the SFnAv, as a phenomenological characteristic: the SFnAv on budgetary revenue is an SFnA, it controles in a linear way the quantitative change of the budgetary revenue and its action is discrete, i.e., by this device, the aim is achieved through a discrete action.

### *5.6. Definition of the non-automatic fiscal stabiliser on budgetary expenditure*

In our opinion, the SFnAc is generated by the simultaneous checking of three sufficient predicates:

1. it is an SFnA;
2. it controles in a linear way the quantitative change of the budgetary expenditure;
3. its action is discrete, i.e., by this device, the aim is achieved through a discrete action.

Therefore, we may draw the following concluding remarks regarding the definition of the SFnAc, as a phenomenological characteristic: the SFnAc an SFnA, it controles in a linear way the quantitative change of the budgetary expenditure and its action is discrete, i.e., by this device, the aim is achieved through a discrete action.

### *5.7. Definition of the automatic fiscal stabiliser*

In our opinion, the SFA is generated by the simultaneous checking of two sufficient predicates:

1. it is an SF;
2. its action is formal implicit, i.e., by this device, the aim is achieved through a non-discretionary, formal implicit action[15].

Therefore, we may draw the following concluding remarks regarding the definition of the SFA, as a phenomenological characteristic: the SFA is an SF and its action is formal implicit, i.e., by this device, the aim is achieved through a non-discretionary, formal implicit action.

### *5.8. Definition of the automatic fiscal stabiliser on budgetary revenue*

In our opinion, the automatic fiscal stabiliser on budgetary revenue (SFAv) is generated by the simultaneous checking of three sufficient predicates:

1. it is an SFA;
2. it controles in a non-linear[16] way the quantitative change of the budgetary revenue;
3. its action is discrete, i.e., by this device, the aim is achieved through a discrete action.

Therefore, we may draw the following concluding remarks regarding the definition of the SFA on budgetary revenue, as a phenomenological characteristic: the SFAv is an SFA, it controls in a non-linear way the quantitative change of the budgetary revenue and its action is discrete, i.e., by this device, the aim is achieved through a discrete action.

### *5.9. Definition of the automatic fiscal stabiliser on budgetary expenditure*

In our opinion, the automatic fiscal stabiliser on budgetary expenditure (SFAc) is generated by the simultaneous checking of three sufficient predicates:

1. it is an SFA;
2. it controls in a non-linear way the quantitative change of the budgetary expenditure;
3. its action is discrete, i.e., by this device, the aim is achieved through a discrete action.

Therefore, we may draw the following concluding remarks regarding the definition of the SFA on budgetary expenditure, as a phenomenological characteristic: the SFAc is an SFA, it controls in a non-linear way the quantitative change of the budgetary expenditures and its action is discrete, i.e., by this device, the aim is achieved through a discrete action.

## 6. Conclusions on non-automatic fiscal stabilisers vs automatic fiscal stabilisers

- discretionary fiscal instrument (SFnA) is activated both following signalling message on the volatility of macroeconomic output (GDP) and outside it;
- non-discretionary fiscal instrument (SFA) is activated by the automatic achieve of a programmed level on volatility of macroeconomic output (GDP);
- discretionary fiscal instrument (SFnA) is normative explicit;
- non-discretionary fiscal instrument (SFA) is normative implicit;
- discretionary fiscal instrument (SFnA) doesn't generate delays between its application and effect;
- non-discretionary fiscal instrument (SFA) generates delays between its application and effect;
- discretionary fiscal instrument on budgetary revenue (SFnAv) controls in a linear way the quantitative change of budgetary revenue, and its action is discrete;
- discretionary fiscal instrument on budgetary expenditure (SFnAc) controls in a linear way the quantitative change of budgetary expenditure, and its action is discrete;
- non-discretionary fiscal instrument on budgetary revenue (SFAv) controls in a non-linear way the quantitative change of budgetary revenue, and action is discrete;
- non-discretionary fiscal instrument on budgetary expenditures (SFAc) controls in a non-linear way the quantitative change of budgetary expenditure, and action is discrete.

## 7. Mathematical conditions as a fiscal stabiliser to be automatic-type

For establishing the mathematical conditions as a fiscal stabiliser to be automatic-type, we propose the following quantitative equilibrium model:

$$\begin{aligned}|PIBa(t) - PIBp(t)| &= K(t) \cdot B(t) ||SBC(t)| - |SBS(t)|| = \\ &= K(t) \cdot B(t) ||Va(t) - Vp(t)| - |Ca(t) - Cp(t)||\end{aligned} \quad (13)$$

where:

PIBa = current GDP;
PIBp = potential GDP;
K = rate of action of SFA;
B = base of action of SFA;
Va = budgetary revenue related to current GDP;
Vp = budgetary revenue related to potential GDP;
Ca = budgetary expenditure related to current GDP;
Cp = budgetary expenditure related to potential GDP;
T = point in time to which they relate all indicators of formula (13), which covers a period of time.

We note the expression $|PIBa(t) - PIBp(t)|$ with *Vol* and call it the volatility function of GDP, so:

$$Vol = Vol(t) = K^{*3}(t) \cdot b^3(t) |N(t) - M(t)| \quad (14)$$

where: $K^* = \sqrt[3]{K}; b = \sqrt[3]{B}; N(t) = |Va(t) - Vp(t)|; M(t) = |Ca(t) - Cp(t)|$.

According to relation (14) we have two possible situations:

A. For $M(t) = N(t)$, it is not necessary to use automatic fiscal stabilisers.

B. For $N(t) \neq M(t)$ it is necessary to use automatic fiscal stabilisers and we have:
   a. for the first derivative:

$$\begin{cases} \dfrac{\partial Vol}{\partial K^*} = \dfrac{\partial Vol}{\partial t} \cdot \dfrac{\partial t}{\partial K^*} = \dfrac{\partial Vol}{\partial t} \cdot \left(\dfrac{\partial K^*}{\partial t}\right)^{-1} = \dfrac{\partial Vol}{\partial t} \cdot 3K^{*2}(t) \\ \dfrac{\partial Vol}{\partial b} = \dfrac{\partial Vol}{\partial t} \cdot \dfrac{\partial t}{\partial b} = \dfrac{\partial Vol}{\partial t} \cdot \left(\dfrac{\partial b}{\partial t}\right)^{-1} = \dfrac{\partial Vol}{\partial t} \cdot 3b^2(t) \end{cases} \quad (15)$$

Points of time, wherein:

$$K^*(t) \to 0, b(t) \to 0 \quad (16)$$

will be solutions of the system of equations (15), where the partial derivatives are equal to 0, so they are stationary points[17] of the volatility function of GDP (14).

   a. for the second derivative:

$$\begin{cases} \dfrac{\partial^2 Vol}{\partial K^{*2}} = \dfrac{\partial^2 Vol}{\partial t^2} \cdot 3K^{*2}(t) + \dfrac{\partial Vol}{\partial t} \cdot 6K^*(t) \\ \dfrac{\partial^2 Vol}{\partial b^2} = \dfrac{\partial^2 Vol}{\partial t^2} \cdot 3b^2(t) + \dfrac{\partial Vol}{\partial t} \cdot 6b(t) \\ \dfrac{\partial^2 Vol}{\partial K^* \partial b} = \dfrac{\partial^2 Vol}{\partial b \partial K^*} = \dfrac{\partial}{\partial K^*}\left(\dfrac{\partial Vol}{\partial b}\right) = \dfrac{\partial^2 Vol}{\partial K^* \partial t} \cdot 3b^2(t) + \dfrac{\partial Vol}{\partial t} \cdot 3\dfrac{\partial^2 b^2(t)}{\partial K^*} \end{cases} \quad (17)$$

[18]As a result,

$$d^2 Vol = \frac{\partial^2 Vol}{\partial K^{*2}} \cdot dK^{*2} + 2\frac{\partial^2 Vol}{\partial K^* \partial b} \cdot dK^* db + \frac{\partial^2 Vol}{\partial b^2} \cdot db^2 \quad (18)$$

$$\frac{\partial^2 Vol}{\partial K^{*2}} \cdot dK^{*2} + 2\frac{\partial^2 Vol}{\partial K^* \partial b} \cdot dK^* db + \frac{\partial^2 Vol}{\partial b^2} \cdot db^2 > 0 \quad (19)$$

in the stationary points, being a sum of positive terms, if we replace in the differential formula of 2nd order, the partial derivatives of the system (17).

We know that the *quantitative dimension of SFA effectiveness* (E) is the product of the rate of action and the base of action:

$$E = -K \cdot B \quad (20)$$

The rate of action and the base of action are interchangeable. In this case, we can define an indifference curve of the effectiveness of SF, with the differential condition (Dinga, 2009, p. 89):

$$dE = 0 \rightarrow dK \cdot B + dB \cdot K = 0 \rightarrow Rms = \frac{dK}{dB} = -\frac{K}{B} \quad (21)$$

At a time moment t for which formula (20) is used, we obtain the quantitative dimension of SFA effectiveness through the analytic expression:

$$E(t) = -K(t)B(t) \quad (22)$$

## 8. Conclusion

As it was shown in numerous scientific works, the economical processes take place in cycles, therefore the evolution of the quantitative dimension of SFA effectiveness has this type of characteristic. Hence, function *E(t)* is the solution of a differential equation as the following:

$$\frac{dE}{dt} = E(1-E) \quad (23)$$

We noted with '1' in formula (23) a 'target' value for the quantitative dimension of SFA effectiveness, which we consider to be the measurement unit on a real axis, on which we represented the values of the quantitative dimension of SFA effectiveness graphically.

The right-hand side of relation (23) is a grade two function with the variable *E*, which has a maximum in the vertex of the parabola. These aspects lead to the conclusion that the variation of *E*, respectively $\frac{dE}{dt}$, is the highest when *E* is equal to the abscissa of the parabola's vertex, which means for $E = \frac{\varepsilon}{2}$.

The inverse proportionality between the rate of action and the base of action results from relation (9), because by separating the variables we obtain:

$$\int \frac{dK}{K} = -\int \frac{dB}{B} \quad (24)$$

By calculating the integrals, we obtain:

$$K(t) = \frac{c}{B(t)} \quad (25)$$

where *c* represents a nonzero real number. This can determine if, at a fixed time moment t, the values for *K(t)* and *B(t)* can be substituted in formula (25).

At the same time, also because of the cyclic evolution of the economic processes, the values of *B(t)* will have an evolution modelled by a logistic equation. This has the form:

$$\frac{dB}{dt} = B(1-B) \qquad (26)$$

and by integrating it, we obtain the following analytic expression for the base of action:

$$B(t) = \frac{e^t}{C+e^t} \qquad (27)$$

The real constant *C* from formula (27) can be determined by substituting the value of *B(t)* at a fixed moment of time *t*.

Therefore,

$$E(t) = -K(t)\frac{e^t}{C+e^t} \qquad (28)$$

and the identification conditions of an optimum *E(t)* are:

$$\frac{dE}{dt} = 0 \qquad (29)$$

(the solution of this equation being a critical point for the evolution of the values of *E*, which will also represent at the same time a equilibrium point) and

$$\frac{d^2E}{dt^2} < 0 \qquad (30)$$

Mathematically, the optimality conditions (29) and (30) are equivalent to:

$$K(t) = e^{-\int \frac{C}{C+e^t} dt} \qquad (31)$$

$$\frac{\left(-CK'(t)e^t - 2CK'(t)e^t - K'(t)e^{2t} - 2K'(t)e^{2t} - K(t)Ce^t\right)(C+e^t)^2 + \left(CK'(t)e^t + K'(t)e^{2t} + K(t)Ce^t\right)2(C+e^t)e^t}{(C+e^t)^4} < 0$$

$$(32)$$

Obviously, formula (32) becomes simpler after substituting *K(t)* with its expression from formula (31).

Observation and methodological precautions:

- mathematical conditions to produce quantitative equilibrium of formula (13) are conditions (16) and (18);
- an SF, to be of automatic-type, it must satisfy conditions (16) and (19);
- the conditions for which E(t) is optimal can be found in relations (31) and (32), for any country and for any time frame;

- as the rate of action and the base action are not continuous nelinear (they are not practically manageable) but discontinuous nelinear, they can be approximated and substituted by nelinear derivable functions, determined by numerical methods (e.g., numerical method of functions interpolation);
- point where Vol(t) = 0 is an attractor;
- the sequence of bifurcations can not be asked for $t \to \infty$, the frequency of returning at some point in the space of phasse being infinite and so, basically, the trajectory no longer runs through any point in the past;
- by analogy with the aggregate model of Hicks of type accelerator-multiplier, on he dynamic equilibrium of GDP evolution, which is reduced to solve equations with finished differences, it can create a model on the evolution of the function Vol(t), function that expresses the deviation absolute of the macroeconomic output at the time (t).

Regarding mathematical background on the methodology for developing and applying mathematical models in formulas (13) and (14), we have used functions that allow the use of differential calculus concepts and theorems in order to determine the optimal points.

In section A, page 12, if *M(t) = N(t)*, we found that the SFAs intervention is not required.

In section B, page 12, in order to draw up the optimal conditions required to find the minimum point of volatility function, we have drafted system (15) for determining stationary points and then system (17) and conditions (18) and (19) to show which of the stationary points are optimal points, and therefore, under what conditions *Vol* function has a minimum.

It is known that the first product in formula (13) underlies the definition of the concept *the Quantitative dimension of SFA Effectiveness* (E) in formula (20) and that it checks the condition in formula (21).

Formula (22) expresses E for an arbitrary moment of time *t*.

After determining *B(t)* in formula (26), one may find the expression of *E(t)* as in formula (28), the optimal point and balance point respectively, being determined according to differential calculation theory in conditions (29), (30), equivalent to (31), (32).

Mathematical models specified in the article may be implemented, validated and verified through the use of specific software, such as Maple 16. As an example, we input equation (26) into the software, to which we added the initial condition *B(2014)=172055.3*. The value selected as an example within the initial condition, represents million lei current prices and was obtained by summing gross wages and salaries in Romania, in 2014, namely: 172055.3 = 39388.9 + 42106.5 + 44315.1 + 46244.8 (values displayed on the Romanian National Institute of Statistics website, www.insse.ro). We considered gross wages and salaries, which are suitable to SFAs (not all revenue is suitable to SFAs – see comment page 7, paragraph 1, of this article).

The software has generated equation (26) solution, as in the following formula:

$$B(t) = \frac{1720553 e^{-2014}}{-1720553 e^{-2014} + 1720543 e^{-t}} \qquad (33)$$

Nowadays and in the future, concerns are directed towards mathematical modelling and simulating economic processes in order to operatively and effectively smooth the economic cycle, to obtain remarkable results. Obviously, given both the complexity of real-world economic processes and mathematical limitations known until now, whatever the mathematical model used or developed, it has an idealised character and provides only part of the problem properties clipped from economy.

## Notes

1. According to the definition, this involves a 'conscious' intervention of the public organisation/institution responsible for that PP.
2. Or explicit fiscal rules, which take the form of Ordinances of the Government (simple or emergency), Decisions of the Government or Laws, in the field of taxation.
3. Or implicit fiscal instruments.
4. Budget balance associated with potential GDP.
5. In the terminology of the IMF, OECD.
6. Cyclically adjusted balance.
7. We will use the scientific classification of the Sciences of nature, biology.
8. We use the concept of category (class in biology) to avoid Russell's paradox. By this category we understand that category attached to the attracting condition close to equilibrium. In thermodynamics, there are three categories of attractions: (a) at equilibrium (equilibrium thermodynamics); (b) close to equilibrium (non-equilibrium linear thermodynamics); and (c) far from equilibrium (nonlinear thermodynamics).
9. Another family of SM is family called monetary stabiliser (SMo).
10. i.e., having a mandatory, normative character.
11. By device, we understand a set of interrelated components to meet an external function / orientated exogenous.
12. By change we understand the meaning of the Aristotle, that change of place, change of quantity, change of quality.
13. By desired change we understand the necessary change (in logical sense).
14. We remember that by the macroeconomic volatility of output (GDP) we understand the overcoming of current GDP by potential GDP (on the increase) or decrease actual GDP below potential GDP (on the decrease).
15. The introduction action of SFA is explicit normative, but its functioning is implicit normative.
16. Regarding the concept of non-linearity, see our work, Bratian, 2012.
17. According to the definition in the field of mathematical analysis.
18. $\frac{\partial^2 Vol}{\partial K^* \partial b} = \frac{\partial^2 Vol}{\partial b \partial K^*}$: Schwartz's theorem (the function is continuous with respect to $K^*$ and b, so the two partial derivatives are equal).

## Disclosure statement

No potential conflict of interest was reported by the authors.

## Funding

This work was supported by Lucian Blaga University of Sibiu [grant number LBUS-IRG-2015-01]

## References

Andres, J., & Domenech, R. (2006). Automatic stabilizers, fiscal rules and macroeconomic stability. *European Economic Review, 50*, 1487–1506. doi:10.1016/j.euroecorev.2005.03.005

Blanchard, O. (1990). Suggestion for a new set of fiscal indicators. *OECD, Economics Department, Working Papers*, 79. doi: 10.1787/435618162862

Bratian, V. (2012). *Economic organization and paradigm of the living logical system*. Saarbrucken: Lambert Academic Publishing

Brătian, V. (2014). Defining the concepts of organization, economic organization and stabilizer from the perspective of complex systems. *Procedia Economics and Finance, 16*, 540–547. doi:10.1016/S2212-5671(14)00836-3

Camba-Mandez, G., & Lamo, A. (2002). Short-term monitoring of fiscal policy discipline. *European Central Bank, Working Paper Series, No. 152*. Retrieved from https://www.ecb.europa.eu/pub/pdf/scpwps/ecbwp152.pdf

Christiano, L., Eichenbaum, M., & Rebelo, S. (2011). When is the government spending multiplier large? *The Journal of Political Economy, 119*, 78–121.
Coenen, G., Erceg, C., Freedman, C., Furceri, D., Kumhof, M., Lalonde, R., … J., & in 't Veld (2012). Effects of fiscal stimulus in structural models. *American Economic Journal: Macroeconomics, 4*, 22–68.
Cogan, J. F., Cwik, T., Taylor, J. B., & Wieland, V. (2010). New keynesian versus old keynesian government spending multipliers. *Journal of Economic Dynamics and Control, 34*, 281–295.
Dalsgaard, T., & De Serres, A. (1999). Estimating prudent budgetary margins for 11 eu countries: A simulated var model approach (OECD, Economics Department Working Papers 216). Retrieved from http://www.oecd.org/eu/1879140.pdf
Davig, T., & Leeper, E. M. (2011). Monetary–fiscal policy interactions and fiscal stimulus. *European Economic Review, 55*, 211–227.
Dinga, E. (2009). *Studii de Economie – contribuții de analiză logică, epistemologică și metodologică* [Economic studies – contributions in logical, epistemological and methodological analysis]. Bucharest: Economic Publishing House.
Dinga, E., (coordinator) (2011). *Sustenabilitatea economică prin politici de ajustare în contextul globalizării* [Economic sustainability through adjusting policies in the context of globalisation]. Bucharest: Romanian Academy Publishing House.
Dolls, M., Fuest, C., & Peichl, A. (2012). Automatic stabilizers and economic crisis: US vs. Europe. *Journal of Public Economics, 96*, 279–294. Retrieved from http://www.izajolp.com/content/1/1/4;
Eichengreen, B. (1997). Saving Europe`s automatic stabilizers. *National Institute Economic Review, 159*, 92–98. Retrieved from http://EconPapers.repec.org/RePEc:sae:niesru:v:159:y:1997:i:1:p:92-98
Fedelino, A., Ivanova, A., & Horton, M. (2009). Computing cyclically adjusted balances and automatic stabilizers. *Tehnical Notes and Manuals, IMF,* November. Retrieved from http://www.imf.org/external/pubs/ft/tnm/2009/tnm0905.pdf
Follette, G., & Lutz, B. (2010). Fiscal policy in the United States: Automatic stabilisers, discretionary fiscal policy actions, and the economy. *Board of Governors of the Federal Reserve System – Finance and Economics Discussion Series*, 2010–43. Retrieved from http://www.federalreserve.gov/pubs/feds/2010/201043/201043pap.pdf
Girouard, N., & Andre, C. (2005). Measuring cyclically-adjusted budget balances for oecd countries (Economics Department Working Papers 434). Retrieved from 10.1787/787626008442
Martner, R. (2000). Automatic fiscal stabilizers. *Cepal Review 70 – Eclac Publicaciones*, 31–55. Retrieved from http://www.eclac.org/publicaciones/xml/2/19912/lcg2095i_Martner.pdf
Suescún, R. (2007). *The Size and effectiveness of automatic fiscal stabilizers in Latin America*. Washington, DC: World Bank. Retrieved from https://openknowledge.worldbank.org/handle/10986/7072
Treaty on stability, coordination and governance, European Council (TSCG – UEM/ro). (2012). Retrieved from http://www.european-council.europa.eu/media/639164/18_-_tscg.ro.12.pdf
Veld, J., Larch, M., & Vandeweyer, M. (2013). Automatic fiscal stabilisers: What they are and what they do. *Open Economies Review, 24*, 147–163. doi:10.1007/s11079-012-9260-6